\documentclass{article}
\usepackage[T1]{fontenc} 
\usepackage[utf8]{inputenc} 
\usepackage{ismir,amsmath,cite,url}
\usepackage{graphicx}
\usepackage{color}
\usepackage{multirow, hhline, array, booktabs, graphics, subfigure}
\usepackage{verbatim}

\title{Towards a New Interface for Music Listening:\\ 
A User Experience Study on YouTube}

\multauthor
{Ahyeon Choi \hspace{1cm} Eunsik Shin \hspace{1cm} Haesun Joung \hspace{1cm} Joongseek Lee \hspace{1cm} Kyogu Lee} 
{ \bfseries
Department of Intelligence and Information, Seoul National University\\
{\tt\small 	{\{chah0623, esshin, gotjs3841, joonlee8, kglee\}}@snu.ac.kr
}
\vspace{-0.3cm}
}

\def\authorname{Ahyeon Choi, Eunsik Shin, Haesun Joung, Joongseek Lee, and Kyogu Lee}

\begin{document}
\maketitle

\begin{abstract}
In light of the enduring success of music streaming services, it is noteworthy that an increasing number of users are positively gravitating toward YouTube as their preferred platform for listening to music. YouTube differs from typical music streaming services in that they provide a diverse range of music-related videos as well as soundtracks. However, despite the increasing popularity of using YouTube as a platform for music consumption, there is still a lack of comprehensive research on this phenomenon. As independent researchers unaffiliated with YouTube, we conducted semi-structured interviews with 27 users who listen to music through YouTube more than three times a week to investigate its usability and interface satisfaction. Our qualitative analysis found that YouTube has five main meanings for users as a music streaming service: 1) exploring musical diversity, 2) sharing unique playlists, 3) providing visual satisfaction, 4) facilitating user interaction, and 5) allowing free and easy access. We also propose wireframes of a video streaming service for better audio-visual music listening in two stages: search and listening. By these wireframes, we offer practical solutions to enhance user satisfaction with YouTube for music listening. These findings have wider implications beyond YouTube and could inform enhancements in other music streaming services as well.

\end{abstract}

\section{Introduction}\label{sec:introduction}

In recent years, the music streaming industry has witnessed a significant surge in popularity, with market leaders such as Spotify, Apple Music, and Amazon Music dominating the market\cite{Global}. Alongside this trend, YouTube has solidified its position as a prominent platform for diverse video content, including documentaries, daily vlogs, entertainment shows, and more. As users flocked to YouTube for various types of content, the platform naturally became a hub for music-related videos as well. Users now have easy access to a wide range of music video content on YouTube, contributing to the growing trend of consuming music through video formats\cite{book}.

Indeed, YouTube delivers a distinctive multi-sensory experience by showcasing a vast variety of music-related videos such as music videos, live performances, curated playlists with visual artworks, and cover performances, enabling users to enjoy music through a fusion of visual and auditory elements. Despite Spotify's global prominence based on subscribers, YouTube has seen an increasing number of users turning to its platform for music consumption\cite{Global, 29}. This trend is evident in regions like South Korea\cite{korea, kocca} and Latin America\cite{Latin}, where YouTube dominates as a preferred music platform.

Given YouTube's current dominance in music consumption, there's a need for a more comprehensive investigation into this behavior and patterns. Earlier studies have explored YouTube's role as a streaming service\cite{28}, compared its usability with Spotify\cite{29}, and analyzed music consumption behavior on YouTube\cite{27}. However, the elements contributing to YouTube's rise as a primary music platform and the actual levels of user satisfaction are still not fully understood, indicating a need for further user-focused research.

Thus, this study aims to conduct in-depth interviews with music consumers on YouTube, examining their behavior, comparing the advantages and disadvantages of using YouTube as a music consumption tool with other music streaming services, and reevaluating YouTube's standing as a tool for music consumption. Additionally, we propose a new interface design that enhances the usability of music-related searches and listening. This research was conducted independently by our team, with no financial backing or data provided by YouTube or any associated organization. With our study, we aim to contribute to the ongoing conversation on YouTube's role as a music platform and offer insights into developing an innovative interface that elevates the user's music listening experience.

\section{Related Work}
In the field of music information retrieval (MIR), research on music streaming services includes studies on improving recommendation algorithms
\cite{Rec1, Rec2, Rec3},
understanding user behavior and patterns of use
\cite{beh1, beh2, beh3, beh4, beh5},
and studying user experiences and interfaces
\cite{exp1, exp2, exp3, exp4, exp5}. These studies aimed to enhance overall user satisfaction and engagement with music streaming services by providing personalized recommendations, improving the user interface, and identifying the factors that influenced user behaviors and preferences.

Compared to other music streaming services, research on music consumption through YouTube has only recently gained attention due to the platform's relatively late recognition as a music consumption platform. Early studies on YouTube's music videos have revealed that music is the most consumed content category on YouTube, and researchers have classified the types of YouTube's music content while analyzing their differences\cite{28}. Furthermore, \cite{29} reported that YouTube is used as frequently as Spotify and is perceived as superior to Spotify in terms of its shareability and accessibility. 

As YouTube's influence in music consumption grows, recent research has examined three types of online music practices according to the role YouTube plays: default, soundtracking, and complementary platforms\cite{27}. Authors report that one of the main results is that YouTube's music videos are listened to, rather than watched. However, the significance of visual elements in music listening can differ based on the genre or content. Additionally, it is worth mentioning that the participants in the study reported only occasional use of YouTube for music, which may limit the generalizability of the findings to other contexts, such as frequent YouTube users.

Therefore, this study aims to examine the usage behavior of users who use YouTube more than three times a week in everyday situations, report on the characteristics of the subject group, and classify the content used. In addition, we draw out advantages and disadvantages through usability tests to newly consider the role of YouTube as a music-listening tool. Moreover, the study proposes interface improvement measures to fill the research gap on "how to improve the music listening environment through YouTube." Considering the diverse range of devices used to access YouTube, including mobile devices, PCs, tablets, and TVs, we primarily focus on the mobile device, taking into account its widespread usage among participants.

\vspace{-0.1cm}
\section{Methods}

\subsection{Participant}
We recruited 27 Seoul National University students (12 males, 15 females) aged 18 or older (mean=23.40, sd=3.13). Our recruitment focused on participants who listen to music on YouTube at least three times a week while excluding those who rely solely on YouTube Music without using YouTube. This approach allowed us to concentrate on the distinct characteristics of consuming music through videos on YouTube, which encompass both visual elements and audio. Participants were compensated with a cash payment of KRW 10,000. Ethics approval was obtained from the Institutional Review Board of SNU.

\subsection{Study Design}

\newcolumntype{A}{>{\centering\arraybackslash}m{1.7cm}}
\newcolumntype{B}{>{\centering\arraybackslash}m{5.3cm}}
\begin{table}[ht]
    \begin{center}
    \begin{tabular}{A|B}
    \hline
    Phase & Requirements\\
    \hhline{=|=}
    A Verbal Interview & Asking about participants' music listening habits and preferences along with the motivation to use YouTube. \\
    \hline
    Usability Test & Comparison of YouTube and other music streaming services and feedback on the interface of YouTube for searching and listening to music. \\ \hline
    UI proposal & Propose YouTube interface design for music listening freely, and explain yourself.  \\ \hline
    \end{tabular}
    \end{center}
\caption{Three steps of semi-structured interview}
\vspace{-0.3cm}
\end{table}

Informed by previous studies' methodologies and the specific needs of our research, we designed our interview in two stages: a preliminary questionnaire\cite{8,bunian2021vins}, followed by a semi-structured interview\cite{8,20,bunian2021vins} that includes a brief ice-breaking session\cite{kindbom2022does}. The preliminary questionnaire collects demographic data and music consumption habits of the participants, such as their academic majors, relationship with music, frequency and duration of YouTube use for music, and specific contexts of YouTube music consumption (excluding YouTube Music). Additionally, we also sought information regarding their subscription to YouTube Premium or usage of YouTube Music.

Following an ice-breaking session, the semi-structured interview proceeded with three main segments (Table 1). First, we explored participants' regular music consumption habits, such as frequency, platform preference, and content preferences. Second, participants were asked to demonstrate the process of searching and listening to music on YouTube, which allowed for a natural exploration of the platform's advantages and disadvantages in comparison to other music streaming services. Third, participants utilized empty interface templates on iPad to design a new interface for music searching and listening, enabling them to customize the screen ratio, functions, buttons, and more. Each interview, lasting roughly 30-40 minutes, was recorded and transcribed using NAVER Clova Note, with participant consent.

\subsection{Analysis}
We identified the overarching themes and trends of the participants' responses and organized the data accordingly. The data were categorized into the following topics: primary streaming service, weekly listening time, music listening type, preferred music genres or content on YouTube, situations YouTube is used for music listening, reasons for using YouTube as a music consumption tool, music search methods on YouTube, criteria for video selection, advantages and disadvantages of YouTube compared to other services, and a summary of interface proposal sessions.

We generated a list of keywords for the qualitative analysis, which includes the advantages and disadvantages of using YouTube for music listening and user interface proposals from interviewees. To validate our classifications and identify commonalities, we repeated the process of analysis and consensus-building three times among the researchers similar to the analysis process in \cite{8, 16, 21}. Grounded theory\cite{30} and content analysis\cite{11-1} were also used as a guide throughout the process of keyword generation. We also referred to previous qualitative studies in the field of MIR\cite{beh4,16,20,21,27} to guide our data analysis, as well as to ensure consistency in our reporting and citation practices. Finally, we thoroughly reviewed the keyword lists to extract the main findings of how YouTube is used as a music consumption tool by the participants based on the method of theme analysis\cite{12}.

To better understand the participants' interface design proposals, we compared the proposals from the participants and reviewed the summary of the interface designing sessions. From this process, we synthesized useful design implications and arrived at wireframe designs for the music search and listening screens.

\section{Result}

\subsection{Behavior and Characteristics of Music Consumption on YouTube}
As the current study investigates interview data from a sample of 27 users, it is important to take into account the unique characteristics of this group. Therefore, information concerning the participants' music consumption behaviors and preferences was gathered through preliminary surveys and interviews. The results showed that participants typically used YouTube about five times per week (mean = 4.89, sd = 1.93), for a total of approximately five hours (mean = 5.35, sd = 3.76), to listen to music while engaging in various activities, such as studying, relaxing, commuting, and exercising. No one specialized in music. The majority of participants used the free version of YouTube and did not subscribe to YouTube Premium. Additionally, some participants supplemented their music listening with other platforms such as YouTube Music, Melon, Spotify, and Genie.

Participants enjoyed a diverse range of music genres on YouTube.  The top five genres mentioned most frequently were OST (original soundtrack of movies or dramas, 13 times), pop (12 times), K-pop (11 times), classical music (7 times), and indie music (7 times). Other genres mentioned in order of frequency include J-pop, ballads, old-fashioned music (mid-20th-century Korean pop and ballads), jazz, rock, band music (with live instrumentation and elements of rock, pop, and indie), new age, hip-hop, EDM, and R\&B. Music content can be broadly divided into three categories: 1) Official music content such as music videos, 2) Live music content such as performances, concerts, festivals, and 3) User-generated content such as playlists and cover videos. In terms of frequency of mention, the order was 3-2-1 (27 times, 25 times, 8 times) respectively.

\subsection{Advantages of using YouTube for music listening}

Alongside our anticipation that YouTube serves as an audiovisual music listening tool, we found that YouTube possesses various strengths compared to other streaming services (Table 2). Musical diversity was the most frequently mentioned category, with two main points: the availability of non-official music in addition to official releases, and the diversity of playlist content compared to other streaming services.\vspace{-0.3cm}\\

\noindent\textit{With streaming services, I can only listen to official releases, \textbf{but with YouTube, I can listen to not only official releases but covers and other user-generated content.}} (P11)\vspace{-0.3cm}\\

\noindent\textit{Unlike other services, \textbf{YouTube's diverse playlists prevent repetitive listening by offering a wide range of songs within similar genres.}} (P26)\vspace{-0.3cm}\\

\newcolumntype{C}{>{\centering\arraybackslash}p{2.4cm}}
\newcolumntype{D}{>{\centering\arraybackslash}p{3.2cm}}
\newcolumntype{E}{>{\centering\arraybackslash}p{0.6cm}}
\begin{table}[t]
    \begin{tabular}{C|D|E|E}
    \hline
    Category & Keyword & Freq & Total \\ \hhline{=|=|=|=}
    \multirow{2}{*}{Musical Diversity} & official soundtrack + $\alpha$ & 14 & \multirow{2}{*}{23} \\ \cline{2-3}
     & playlist & 9 &  \\ \hline
    \multirow{4}{*}{Convenience} & familiarity & 4 & \multirow{4}{*}{15} \\ \cline{2-3}
     & accessibility & 4 &  \\ \cline{2-3}
     & subscription fee & 4 &  \\ \cline{2-3}
     & Customizing & 3 &  \\ \hline
    \multirow{2}{*}{User Interaction} & recommendation & 10 & \multirow{2}{*}{13} \\ \cline{2-3}
     & comments & 3 &  \\ \hline
    \multirow{2}{*}{Visual Contents} & thumbnail & 6 & \multirow{2}{*}{10} \\ \cline{2-3}
     & video & 4 &  \\ \hline
    etc. & etc. & 1 & 1 \\ \hline
    \end{tabular}
\caption{Pros. keywords of usability test}
\vspace{-0.3cm}
\end{table}

Also, convenience was mentioned as an advantage, with familiarity, accessibility, no subscription fee, and user customization.\vspace{-0.3cm}\\

\noindent\textit{I use it because I'm used to it. I've used Melon and YouTube Music before, \textbf{but I settled with YouTube because it was more convenient.}} (P12)\vspace{-0.3cm}\\

\noindent\textit{\textbf{Since YouTube is free, there's no need to pay for other services.}} (P8)\vspace{-0.3cm}\\


As for user interaction, most users mentioned the recommendation algorithm itself and the ability to view other users' opinions through comments.\vspace{-0.3cm}\\

\noindent\textit{\textbf{The recommendation algorithm is good.} I often find great new songs through it.} (P17)\vspace{-0.3cm}\\

\noindent\textit{\textbf{It's good to be able to see other people's opinions and sympathize by reading comments.}} (P4)\vspace{-0.3cm}\\

Lastly, the visual content of thumbnails and videos was mentioned as an advantage.\vspace{-0.3cm}\\

\noindent\textit{\textbf{I can use both sight and sound when listening to music with videos.}} (P4)\vspace{-0.3cm}\\

\noindent\textit{\textbf{When I play playlists with thumbnails, like at a housewarming party, it adds to the atmosphere, and it's good for interior purposes too.}} (P6)\vspace{-0.3cm}\\

While we initially expected the inclusion of visual elements to be a significant advantage of YouTube, the participants' usage patterns proved more diverse. Some appreciated the visual components, while others turned to YouTube strictly for audio during activities like work or sleep (P2, P5, P22, P23, P24). These observations align with prior research\cite{27}, showing the varied ways users utilize YouTube for music. Although some mentioned listening to audio with the screen off (P7, P23, P24), we excluded this aspect from our analysis as it's a feature exclusive to YouTube Premium subscribers.

\subsection{Disadvantages of using YouTube for music listening}
The inconveniences and disadvantages of listening to music on YouTube were categorized into seven major themes (Table 3). The most frequently mentioned inconvenience was related to user interaction, with many complaints about the inconvenience of filtering the desired information while exploring recommended videos and comments.\vspace{-0.3cm}\\

\newcolumntype{F}{>{\centering\arraybackslash}p{2.5cm}}
\newcolumntype{G}{>{\centering\arraybackslash}p{0.55cm}}
\begin{table}[t]
    \begin{tabular}{F|F|G|G}
    \hline
    Category & Keyword & Freq & Total \\ \hhline{=|=|=|=}
    \multirow{2}{*}{User Interaction} & comments & 12 & \multirow{2}{*}{21} \\ \cline{2-3}
     & recommendation & 9 &  \\ \hline
    \multirow{2}{*}{Manipulation} & button / tap & 10 & \multirow{2}{*}{17} \\ \cline{2-3}
     & display ratio & 7 &  \\ \hline
    \multirow{3}{*}{Playlist} & playlist contents & 4 & \multirow{3}{*}{10} \\ \cline{2-3}
     & making playlist & 4 &  \\ \cline{2-3}
     & mix playlist & 2 &  \\ \hline
    \multirow{2}{*}{Section Search} & timestamp & 6 & \multirow{2}{*}{9} \\ \cline{2-3}
     & playback bar & 3 &  \\ \hline
    \multirow{2}{*}{Lack of Info.} & song information & 5 & \multirow{2}{*}{8} \\ \cline{2-3}
     & log information & 3 &  \\ \hline
    \multirow{2}{*}{Underutilization} & replay & 5 & \multirow{2}{*}{8} \\ \cline{2-3}
     & volume control & 3 &  \\ \hline
    \multirow{2}{*}{Contents Quality} & sound quality & 5 & \multirow{2}{*}{6} \\ \cline{2-3}
     & video quality & 1 &  \\ \hline
    \multirow{2}{*}{etc.} & (video) data size & 4 & \multirow{2}{*}{6} \\ \cline{2-3}
     & etc. & 2 & \\ \hline
    \end{tabular}
    \caption{Cons. keywords of usability test}
    \vspace{-0.3cm}
\end{table}

\noindent\textit{In other music streaming services, genre separation is clearly done, \textbf{but YouTube recommends based on the videos you watch, so there is a tendency to lean towards a specific genre.}} (P26)\vspace{-0.3cm}\\


\noindent\textit{\textbf{When watching music videos and reading comments, it's hard to find South Korean users' reactions when most comments are in foreign languages.}} (P11)\vspace{-0.3cm}\\

The second most frequently mentioned disadvantage was related to screen manipulation, such as fixed thumbnails, the ratio of videos, and accidental button presses.\vspace{-0.3cm}\\

\noindent\textit{\textbf{It would be nice if I could reduce the screen ratio.} I want to watch the small screen when exercising or doing other things.} (P8)\vspace{-0.3cm}\\

\noindent\textit{\textbf{There are cases where I accidentally press the Shorts button and the music stops.}} (P22)\vspace{-0.3cm}\\

Regarding playlists, users complained about not having timestamps for individual songs, the content of playlists made by others, the process of creating playlists themselves, and the mixes provided by YouTube.\vspace{-0.3cm}\\

\noindent\textit{It's inconvenient to switch to another song \textbf{if there is no timestamp in the playlist.}} (P18)\vspace{-0.3cm}\\

\noindent\textit{\textbf{Since playlists are made by others, there are few cases where all songs suit my taste,} and there are mediocre songs in between.} (P18)\vspace{-0.3cm}\\

\noindent\textit{\textbf{It's inconvenient to save songs one by one in my library.} It feels slow every time I press the save button, and it is a hassle to press the button several times to save.} (P27)\vspace{-0.3cm}\\


Some users mentioned the lack of information about album or song information and lyrics, as well as the lack of log information about previously watched videos as a disadvantage.\vspace{-0.3cm}\\

\noindent\textit{\textbf{It's hard to find album or song information,} and it's frustrating not knowing the information of the concert I am watching.} (P7)\vspace{-0.3cm}\\

\noindent\textit{\textbf{When I use the autoplay function, it is hard to find which song I thought I liked.}} (P17)\vspace{-0.3cm}\\

Despite the existence of autoplay and volume control features on YouTube, user complaints arose from a lack of information about these functions. Some users viewed them as drawbacks, unaware of their existence or location. Specifically, enabling autoplay requires navigating to the settings, while fine-tuning volume necessitates physical device button use. This complexity may have heightened user frustration and dissatisfaction.\vspace{-0.3cm}\\

\noindent\textit{\textbf{I wish there is a autoplay button.}} (P12)\vspace{-0.3cm}\\

\noindent\textit{I want to make minute adjustments, \textbf{but even if I increase the volume level by just one, the volume suddenly becomes too loud.}} (P16)\vspace{-0.3cm}\\

The quality of the content is related to the audio or video quality. Some responses showed low reliability in audio quality when used for music listening.\vspace{-0.3cm}\\

\noindent\textit{There are cases where the \textbf{sound quality is poor in content uploaded by individual users.}} (P12)\vspace{-0.3cm}\\

Aside from that, there were four mentions of concerns about mobile data usage due to large video data size (P2, P9, P11, P25), one mention of discomfort with provocative titles (P23) and experiencing an error when randomly playing saved videos (P14). There were nine mentions related to ads or background playback (P2, P5, P6, P8, P9, P12, P17, P19, P25), but these were excluded from the analysis since they can be resolved with a YouTube premium subscription.


\subsection{User Feedback for Interface Improvements}
We analyzed users' explanations and drawings of the searching and listening screens, categorizing their demands for interface improvement into three categories: addition, modification, and deletion. These categories, along with relevant quotes, provide specific descriptions of users' interface improvement suggestions.

The first is to request the addition of new information or functions that are currently absent on YouTube, such as a new button or tab, new sorting and filtering criteria, or more information about contents and songs.\vspace{-0.3cm}\\

\noindent\textit{\textbf{It would be great if there was a detailed search button under the search bar}, where you could search by year, album, composer, etc.} (P7)\vspace{-0.3cm}\\

\noindent\textit{\textbf{It would be nice if I can choose options between "all videos recommended" and "music-related recommendations" in the recommendation section.}} (P9)\vspace{-0.3cm}\\



The second is to modify the existing functions or configuration of YouTube to increase operability and efficiency when searching and listening to music, such as changing the ratio of various spaces on the interface or changing the positions of existing buttons and information.\vspace{-0.3cm}\\

\noindent\textit{It would be great \textbf{if the thumbnail (album cover) could be smaller,} and the title, artist, etc. could be displayed next to it.} (P12)\vspace{-0.3cm}\\

\noindent\textit{\textbf{It would be nice to adjust the ratio of the comment box and recommended videos} so that you can view them together.} (P4)\vspace{-0.3cm}\\

Lastly, there were cases where demands were made to remove things from the existing YouTube interface that are not directly related to music searching or listening.\vspace{-0.3cm}\\

\noindent\textit{\textbf{We don't need the buttons for uploading videos on the bottom menu bar.} It would be great if we could freely configure this menu bar.} (P15)\vspace{-0.3cm}\\

\noindent\textit{\textbf{If we could hide the buttons we don't use often and press the detailed button to show them, it would be neat.}} (P19)\vspace{-0.3cm}\\

\section{Findings and Dicsussion}

\begin{figure*}[t]
    \centering
        \subfigure[Searching]{\label{fig:searching}\includegraphics[width=\columnwidth]{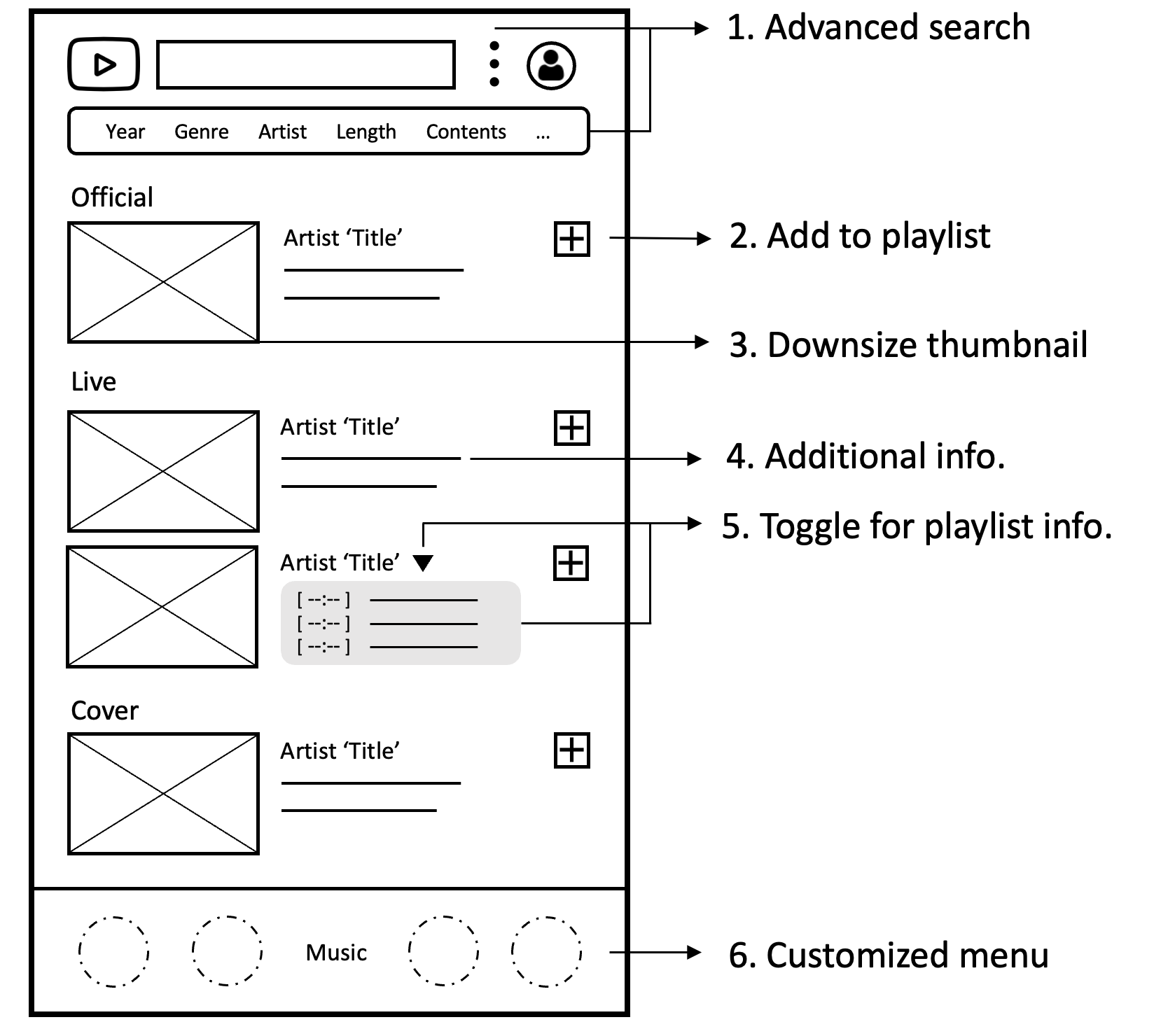}}
        \subfigure[Listening]{\label{fig:viewing}\includegraphics[width=\columnwidth]{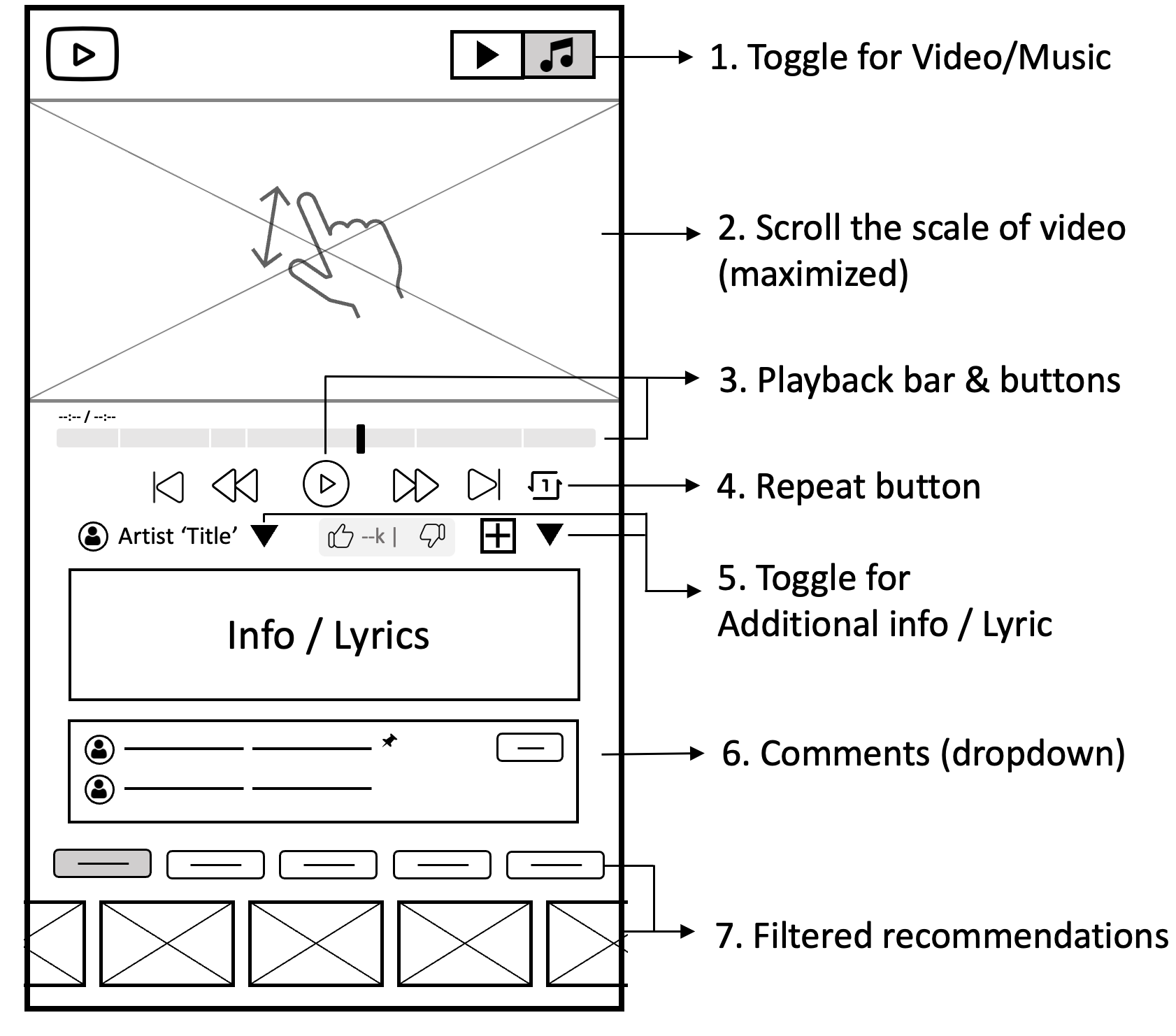}}
        \caption{A Wireframe of youtube UI for music listening}
    \vspace{-0.3cm}
\end{figure*}

\subsection{Role of YouTube as a music streaming service}
Users desire an improved interface for YouTube to maximize its potential as a music consumption tool. We have identified five key roles that YouTube plays in music listening, and based on this, we propose design implications to enhance the user experience.

\subsubsection{Exploring musical diversity}
YouTube offers users a wide variety of musical genres, artists, and songs to discover and explore, including rare or unreleased music not found on other streaming services. Users can also enjoy various versions of the same song through covers or live performances by different artists.
\textbf{Design Implication:} To improve search efficiency, music content should be categorized by genre, artist, and mood, and album information such as lyrics should be provided to reduce the need to search for information on other platforms.

\subsubsection{Sharing unique playlists}
YouTube creators can create and share playlists, simplifying the search process and enabling them to select playlists based on keywords like mood or activity (e.g. warm spring day, driving playlist).
\textbf{Design Implication:} Playlists should provide song and timeline information, and allow users to switch to the next song with a button. Allowing users to customize songs within the playlist, such as adding or removing them, and saving these changes, would enhance the playlist's functionality.

\subsubsection{Providing visual satisfaction}
By offering sensory satisfaction beyond just videos, YouTube's visual content enhances the music listening experience. Users appreciate being able to observe the musicians' expressions, gestures, and style, and sometimes even watch music videos solely for visual gratification like repetitive animations or thumbnail images paired with the music. 
\textbf{Design Implication:} The screen size and ratio of the video should be customizable based on the content's characteristics and users' listening environment. For example, users would like the option to decrease the video screen size in public settings or enlarge it to focus on a particular idol member or musician's finger movements.

\subsubsection{Facilitating user interaction}
YouTube's likes, dislikes, subscriptions, and comments features enable users to interact with the platform and foster a sense of community, resulting in a more engaging music listening experience. Additionally, the recommendation algorithm lets users explore new content and see how others react to music, which is a key motivation for users to use YouTube.
\textbf{Design Implication:} Users should be able to sort and filter recommendations and comments based on various criteria, such as timeline, keyword, lyrics or the most frequently mentioned, to expand YouTube's social function. Pinning specific comments that users like or refer to frequently could also reduce search time.

\subsubsection{Allowing free and easy access}
YouTube's accessibility, cost-effectiveness, and cross-device compatibility make it a convenient option for users to listen to music in various situations. This versatility has led some users to cease subscribing to other streaming services. Primarily, the appeal lies in the free access to a diverse library of music videos, live performances, covers, and user-generated content, resonating with users disinclined to pay for music subscriptions.
\textbf{Design Implication:} While device-specific interfaces are important, consistent usability is crucial to prevent user confusion or inconvenience.

It is crucial to acknowledge that while some of the proposed features (e.g., album and artist filters, lyrics, smaller screen mode) have already been implemented in the YouTube Music App, users still rely on YouTube to access a diverse range of music videos that are not available on the YouTube Music App. Therefore, our design implications hold the potential to differentiate YouTube from YouTube Music by catering to the experience of video streaming alongside music consumption.

\subsection{UI for Audio-Visual Music Streaming Platform}

Taking into account the role of YouTube as a music streaming service, the needs of its users, and the interface designs of typical music streaming services, we have developed an ideal wireframe for an audio-visual music listening platform. It consists of two stages: (a) searching and (b) listening screens (Figure 1).

\textbf{(a) Searching} To display diverse music content tailored to users' interests, we added 1) advanced search functionality to the top keyword search bar, allowing users to filter by era, genre, artist, and other details. Additionally, we added 2) a button to easily add multiple videos to a user's playlist, and 3) reduced the thumbnail size to show more videos on one screen. Next to the thumbnail, we included 4) information about the songs in the video, and if the video is a playlist, we added 5) a timeline and information about the included songs. Finally, we made 6) the bottom menu buttons customizable, allowing users to remove buttons when they feel unnecessary and create their own menu.

\textbf{(b) Listening} While maintaining the current structure of the interface, we adjusted the layout and added new features to enhance the music listening experience. 1) Adding a toggle button that allows users to exchange between video watching and music listening. Users can use their fingers to 2) zoom in or out of the video to adjust its size. Previously, users had to click the video to access playback and skip buttons, but we located 3) the playback bar and related functions at the bottom of the video. We also made the 4) repeat button more visible. We added 5) a toggle button to expand or shorten album information or lyrics, and made 6) comments expandable in a similar manner, with a function for users to pin comments they want to keep visible. We added 7)a filtered recommendation feature to suggest reduced-size videos based on specific user-selected filters. This allows for easier exploration of related content through horizontal scrolling.

The findings of this research hold potential for application across a variety of streaming services. Features such as advanced search functions, customizable menus, and enhanced playlist capabilities can improve user engagement and satisfaction. Effective presentation of music-related information enriches the listening experience, while additional functionalities such as video zoom or comment pinning foster a personalized user experience. These findings can significantly benefit YouTube, as well as aid other music streaming platforms like Spotify, Apple Music, and Amazon Music, and video streaming services including music videos, like Bilibili and Vimeo, in optimizing their interfaces according to their unique characteristics and users' needs.

\vspace{-0.2cm}
\section{Conclusion}
This study explored the music listening behaviors of YouTube users and analyzed the advantages and disadvantages of YouTube as a music streaming service. We proposed new interface wireframes to improve usability and re-examined YouTube's role as a tool for music listening. Undoubtedly, there are constraints in actualizing the proposed interface fully on YouTube. Nevertheless, some suggestions on improving visual satisfaction, comment exploration, and toggle button to exchange the interface between video watching and music listening mode could be considered in designing the overall interface of video streaming platforms.

Our study has limitations depending on the small sample size of Korean users, and it is essential to consider several important factors. Firstly, our interface design primarily focused on mobile environments, which may limit its direct applicability to other devices like PCs and TVs. Secondly, the relatively narrow age range and educational levels of our participants may affect the generalizability of our findings. Thirdly, the absence of comparative studies on similar video platforms and services hinders our understanding of YouTube's performance as a video streaming service. However, these limitations present opportunities for future research to explore and address the diverse needs of users across different devices, demographics, and video services. Overall, our study provides valuable insights and paves the way for further advancements in user-centered design for music streaming services.

\bibliography{YoutubeUI}

\end{document}